# Exploring Faculty Identity Sharing: A Pathway to Empathy in Physics Faculty

Alia Hamdan[1,2]*, Ash Bista[1,] Dina Newman[2], & Scott Franklin[1*]

[1]*Department of Physics, Rochester Institute of Technology, Rochester, USA;*

[2]*Department of Life Sciences, Rochester Institute of Technology, Rochester, USA*

Corresponding author(s). Email(s): ajhcos@rit.edu; svfsps@rit.edu.

## Abstract

**Purpose/Background:** This study investigates how faculty acquire contextual information about students, examining mechanisms and motivations used when sharing their identity to facilitate empathy. Research indicates that unintended behaviours/thoughts of faculty can cause real harm to students (Dancy & Hodari, 2023). Empathy, a multifaceted construct, defined as "the ability and tendency to share and understand others' internal state" (Zaki & Ochsner, 2012**)**, is a critical factor in both motivating faculty to enact large-scale change and take immediate, smaller actions.  Studies on empathetic processes distinguish between cognitive and emotional (affective) empathy and emphasize the significance of gathering contextual information as an initial step toward empathy (Yu & Chou, 2018). The study also explores the impact identity sharing has on obtaining contextual information that motivates empathetic action.

**Methods:** Nineteen semi-structured interviews with physics faculty explored participant identities and interactions with students and colleagues across various contexts. Utilising the empathetic framework, we concentrated on the initial stage of the empathetic pathway—connecting noticing to empathy (Yu & Chou, 2018). Employing emergent thematic coding, we crafted personas around faculty sharing, teaching values, and student reciprocity.

**Results:** We identified four personas. Brooke, the Trust Builder, prioritizes creating an environment of trust by openly discussing their identity, aiming to foster student openness. Wray, adopting a Walled-Off approach, separates personal and professional life due to past negative experiences or a belief in the importance of that division. Casey, the Cautious Sharer, expresses concerns about potential alienation or backlash, approaching personal sharing with caution.  Lastly, Nour, the Identity Navigator, shares personal experiences to assist others in navigating their own identities, acknowledging the challenges of college years. Brooke and Nour had more students approaching them with personal issues, indicating a correlation between faculty identity sharing and student openness. Among the faculty interviewed, 80% who were explicitly open about their identities reported that students approached them with personal issues.

**Conclusion:** Though faculty generally expressed concern for students, not all had the opportunity to engage empathetically. This study outlines mechanisms influencing when and what faculty share about themselves in different contexts. Our findings underscore the significance of fostering dialogue as the initial step in empathy development.

# 1    Introduction

*If we want to even have empathy for something, we have to know if there is a problem, and we should be able to put ourselves in other people's shoes. ~ Dylan [faculty participant]*

Faculty members such as Dylan often aim to connect with their students. Being a professor involves teaching and helping others, and most faculty members genuinely want to do a good job. Empathy is key for building connections and understanding how to support others. However, developing empathy is not a trivial task and requires gaining information about others. Some people are perceptive and can pick up on cues, while most others rely on open communication. Unfortunately, students may find it challenging to share setbacks with their professors, creating a barrier to receiving support. This study emerged from this dilemma, adopting a faculty-centred approach to explore how and when faculty can encourage students to confide in them, thereby taking the initial steps toward building meaningful connections.

This study acknowledges the significant role of faculty members as catalysts for change at both local and broader departmental and institutional levels. Our motivation lies in comprehending the emotional states of faculty members during their interactions with students and their decision-making processes. Empathy is explored as a pivotal factor in student-faculty interactions, serving as a guiding framework throughout this paper. In the following sections, we delve into the importance of empathy in understanding faculty. Subsequently, we transition to an examination of the current state of empathy research, elucidate each author's positionality and involvement in the study, and lastly present the research questions.

## *1.1  The construct of Empathy*

Empathy is a multifaceted construct with research in fields spanning from psychology to education. Due to the multidisciplinary nature of work on empathy, many claim there is not a clear consensus on how empathy can be operationalized (Coplan, 2011; Engelen & Röttger-Rössler, 2012). For our purposes we define empathy as "the ability and tendency to share and understand others' internal state" (Zaki & Ochsner, 2012). Research has recognized two dimensions of empathy: cognitive and affective/emotional. Cognitive empathy involves intellectually grasping another person's perspective or emotions (Gladstein, 1983), while affective or emotional empathy entails emotionally connecting with the other person's experience or feelings (Reniers et al., 2011).

Most work on empathy looks to define the characteristics of the constructs and the pathways through which it develops. Eklund and Meranius (2021) identify four essential traits for an individual to practise empathy: understanding, feeling, sharing another person's world, and self-differentiating from the recipient of empathy. Neurobiologists Yu and Chou (2018) contribute a neurological framework, describing emotional empathy as a reflexive neurological response occurring almost instantaneously, in contrast to cognitive empathy, which they note requires a conscious intellectual effort.

Examining empathy as experienced by physics faculty can illuminate its impact on interactions within academia, influencing faculty perspectives on their roles and shaping future interactions. The importance of empathy in a learning space has long been acknowledged, contributing to the establishment of strong relationships and fostering moral development (Lunn et al., 2022). Instructors with high empathy toward students demonstrate improved impacts on student achievement (Postolache, 2020) and garner more respect from students (Feshbach & Feshbach, 2009). Faculty members recognize empathy as crucial, and research shows that faculty empathy develops throughout their careers as they engage with students and become acquainted with the challenges they face (Lunn et al., 2022).

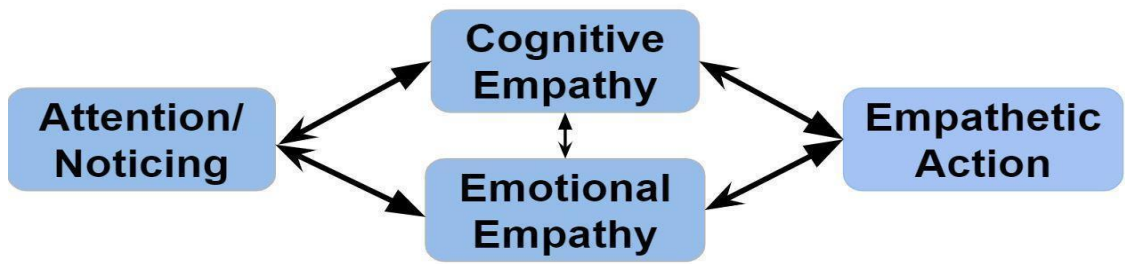

**Figure 1.** Framework for empathetic pathways developed in (Merrill et al, 2024).This image is a simplified illustration of the empathetic pathway, which is separated into two distinct parts. First noticing or paying attention is necessary in order to develop empathy (cognitive and emotional). After empathy is developed, empathetic actions can be taken. Mediators and moderators, such as contextual information, shared lived experiences, are not shown in this figure. It should also be noted that dual arrows are used to indicate the bi-directional nature of these constructs.

Prior work (Merrill et al., 2024) outlined a comprehensive framework delineating empathetic pathways, depicted in Figure 1. This model comprises two primary segments: the first represents the progression from attention and observation to the cultivation of empathy, whether cognitive or emotional; the second depicts the role of empathy in driving action. Merrill et al. additionally found that contextual information mediates the first segment and moderates the second. This paper investigates mechanisms for obtaining contextual information that mediates the development of cognitive and/or emotional empathy.

### 1.2 Communication

The empathy framework often neglects the crucial impact that communication plays in forming the foundation of empathy's development. Communication, as a social process, serves to express thoughts, beliefs, describe events, connect with others, and more broadly, develop social meaning that shapes our understanding of the world (Narula, 2006). While research on communication lacks a consensus on its definition, various models focus on the communication process (Betts, 2009). Three frequently discussed models are the linear, interactive, and transactional models.

The linear model, also known as the Shannon-Weaver model, is the most basic. It involves a sender, a receiver, and a message transmitted through a channel to relay information between two individuals, accounting for the possibility of noise in the channel. In this engineering-oriented approach, the receiver plays a more passive role, and successful communication occurs when the message is transmitted (Cobley, 2013).

Interactive models recognize that communication involves two parties, each acting as a sender and a receiver, creating a communication loop with physical and psychological noise interfering. Physical noise includes environmental factors, such as communication in a loud space, while psychological noise involves the encoding and decoding of messages by the receiver (Cobley, 2013). Effective communication in this model is the creation of shared meaning as individuals' alternate roles.

Transactional models view communication as a process of constructing realities through continuous discourse embedded in social, relational, and cultural contexts. They encompass the interactive model but recognize that all steps occur instantaneously, with messages filtered through participants' experiences. The model considers how each participant uses the shared meaning developed during communication (Cobley, 2013). Despite its nuanced approach, the transactional model relies on active engagement and a deep understanding of nonverbal, cultural messaging, which may or may not be well understood by participants. Miscommunication arises when messages are not interpreted and used as constructed within the feedback loop. Expanding on these models, we explore how faculty utilise communication to experience and apply empathy while engaging with students.

### 1.3     Research Questions

This paper addresses the following research questions:

(1) How do physics faculty approach communicating identity or personal information with colleagues or students in the context of building empathy?

(2) What patterns emerge in physics faculty members' behaviours when discussing or sharing personal information with students and colleagues?

## 2     Positionality statements

Alia Hamdan (she/her) identifies as a Muslim-Palestinian American cis-woman. Alia has navigated higher education physics departments for almost a decade in varying capacities, first as an undergraduate student from a local public university for 4 years, then as a graduate student for a large R1 institute and now as a postdoc at a private R2 institute with a technology focus. Alia was the student who would never go to office hours and often felt out of place in the larger physics culture, but was motivated by her curiosity about the content. Throughout her career, she has sought to make physics a more equitable environment where people can bring their whole selves while engaging in the sciences. She has been on the student and instructor side of navigating physics classrooms and has often felt frustrated and confused about the best ways to move forward/ take action when helping students. Being a visible Muslim creates a lens that others (colleagues, and participants) view her through. Her intersectionality has placed her on the receiving end, or observer of several microaggressions within the field. Alia participated in all parts of this project including the design of the interview protocol, collecting of data, analysis, and writing up the results. At the time of this study, Alia was new to the department and did not know any of the participants before taking part in this study.

Ash Bista (they/her) is an undergraduate physics major who identifies as a first-generation Nepali-American, queer/bisexual, nonbinary/woman of colour from a middle-class background and Hindu family. Ash's experiences centre around the current academic cultures at their institution. Throughout her adolescence and higher education, Ash has struggled with mental health disorders. Despite this, they are very involved in their community and strive to create a positive environment in their classes. They are on the executive board of two student groups, including one that aims to make STEM spaces more inclusive for the LGBTQ community, and work as a teaching assistant. Often, they have volunteered as a peer mentor, panellist, and science outreach demonstrator. Ash knew many of the study participants, having taken classes or worked with them. Ash's intersectional identity, mental health challenges, pre-existing relationship with participants, and personal values contribute to their interaction with the project, shaping how they develop interview protocols, conduct interviews, and interpret results.

Dina Newman (she/her) is the only non-physicist of the group. She identifies as a white woman of Jewish heritage who has been married for over 30 years and is mother to a college-age daughter and a teenage son. As a biology professor and an outsider to the department, she has a unique perspective on the project -- biology has a very different culture from physics, and she has fewer connections to the participants. Dina graduated from an Ivy League university and then went directly to graduate school at another high-prestige R1 institution, which certainly influenced her view of academia. Dina switched her scholarship to biology education research (BER) pre-tenure, about 14 years ago; her research interests are quite broad and reflect her curious and collaborative nature. She has been a feminist since childhood, who has recognized sexism, anti-semitism and elitism in her own environments. Over time she has become more aware of racism and other "isms", leading to her becoming involved in DEIJ activities and incorporating more inclusive practices into her own teaching and research mentoring.

Scott Franklin (he/him) is a white, Jewish, cis-gendered male professor of physics. His upbringing in an Orthodox Jewish day school developed a strong identification with Jewish diaspora and other-ness. An early-diagnosed and medicated neurodiversity (ADHD) provided another experience with feelings of difference. The intersections of Scott's privileged and othered identities shape what he notices and drive his efforts to foster a more inclusive, equitable and socially just STEM and physics environment. Scott received his Ph.D. in a traditional physics discipline but moved into physics education research (PER) during his postdoctoral work. He has been actively involved in the PER community for over twenty-five years, and has been widely recognized for his community-building efforts.

## 3 Methods

### *3.1 Participants*

Nineteen physics faculty from an R2 private university with an enrollment of about 17,000 undergraduates and graduate students were interviewed as part of this study. The department has 35 faculty members, including lecturers and tenure-track positions. One third (32%) of our sample were female, in line with the department average, and half (52%) were people of colour. Seven participants were non tenure-track lecturers with no research component to their jobs. Demographic data can be seen in Table 1. Gender is presented as a binary only because no participants chose to identify otherwise.

Table 1. Participant demographic information

| Demographic Information[a] | | Percentage of Participants[b] |
|---|---|---|
| Faculty Position | Lecturer | 22% |
| | Senior/Principal Lecturer | 16% |
| | Associate Professor | 32% |
| | Assistant Professor | 10% |
| | Professor | 20% |
| Gender | Male | 68% |
| | Female | 32% |
| Ethnicity | Historically marginalised | 52% |
| | White | 48% |

[a]Participant demographics match those of the physics department from which this data was collected.
[b]All percentages were rounded to the nearest 2% range.

### 3.2   *Data Collection*

Participants were recruited through e-mails and announcements at faculty meetings to participate in a series of semi-structured interviews; interviews were conducted by the first or second author during the 2023-24 academic year. Interview protocols for two phenomenographic, semi-structured interviews were designed, tested and modified through an iterative process. The first interview was broad and tackled topics such as identity, conceptual understanding of empathy, how empathy manifests in classroom and research settings, students' struggles in the classroom, and actions taken. This interview developed rapport with the faculty members as well as learned about interactions on a general scope. The second interview was more focused and used instances mentioned in the first interview to more closely investigate thoughts, communication, and actions when dealing with student struggles. Each interview lasted 30-60 minutes. Nineteen faculty participated in the first interview and seven in the second. After the interviews participants were sent a copy of the transcript and given the opportunity to modify (remove, edit) any part of the conversation. (No participants suggested edits or modifications). Pseudonyms were generated at random to maintain anonymity; any gender association with a particular pseudonym is accidental and not affiliated with the gender of the participant. Participants quoted were also sent a draft of this article to confirm the accuracy of both words and meaning (member-checking). One participant chose to elaborate further on their thoughts and the draft includes the new meaningful information.

### 3.3     *Data Analysis*

Data analysis was guided by emergent coding and a constructivist grounded theory approach (Mills et al., 2006), in which the first two authors independently created memos immediately following each interview and again after the initial analysis. We adopted a dual-level approach, beginning with a vertical analysis of summaries of individual participants followed by a horizontal analysis to identify themes across participants. We represent our findings using personas, (Huynh et al., 2021) showcasing themes through fictional characters that embody common characteristics and experiences in a humanlike manner that aligns with the multifaceted aspects of identity (Avraamidou, 2019; Calabrese Barton et al., 2013; Carlone & Johnson, 2007).

The process of creating personas involves several steps. First, we identify and create skeletons—simplified representations of individuals based on coded data. We evaluate and prioritise these skeletons, guided by the framework of Pruitt and Adlin (2006), and decide which skeletons can be developed into personas. Personas are filled in with details derived from coding summaries, capturing commonalities among participants. Two authors (A.H. and A.B.) independently coded participant quotes for the four personas and compared results to identify inconsistencies. Initially, 70% of the quotes matched, with all but one mismatch attributed to variations between the Brooke and Nour personas. These differences were resolved through discussion and clarification about the characteristics of each persona. One excerpt remained unmatched due to the overlapping nature of the personas; dual coding was applied to that excerpt.

*Persona development*

We identified two themes that illustrated instructors' motivation for communication and action and structured our persona-generating analysis on these themes.

(1) Identity-sharing habits in the classroom, research, and with colleagues

We identified instances when identity was shared with students in different settings (classroom, research groups, informal settings, etc.). Participants were asked to identify and speak about different parts of their identity, and we coded when and why those were disclosed. Identity sharing ranged from talking about their academic background to personal examples of when they experienced hardship, such as facing discrimination or bullying in academic spaces. We noted in particular with whom the information was shared and in what context, noting that some chose not to bring to light their identities while others shared often, regardless of the setting.

(2) Teaching values and expectations of role as faculty

The second theme was related to participant's teaching values. Some participants valued clear content delivery, while others emphasised building meaningful connections. Eighteen different codes were identified and divided into four categories as follows:

1. *Growth mindset* codes included teaching students to develop a growth mindset, such as trial and error, or the idea that struggle is part of problem solving.

2. *Safe teaching spaces* codes focused on creating a safe inclusive learning environment, such as being compassionate or students being comfortable to ask questions.

3. *Teaching practices* codes focused on specific teaching practices that the participants found important, like the clarity of information or being well prepared for class.

4. *Socio-cultural constructs* codes aimed at making students think critically of their roles in society at large, such as teaching students to think critically about their future jobs and their influence on society.

# 4 Results

## *4.1 Who communicates first?*

Gaining contextual information is instrumental to cultivating empathy. We find that faculty often struggle to notice non-academic issues that students face and thus do not obtain the contextual information needed to even begin working towards building empathy. Contextual information acts as a moderator in the empathetic pathways and is needed to invoke both or either cognitive and affective empathy (Merrill et al., 2024). Faculty often rely on students to come to them and communicate their needs to work towards a resolution. Some faculty members do mention being proactive about reaching out to students if they notice mainly academic changes.

> "So sometimes I'm perceptive enough that I can try to keep track of most of my students. And, you know, if someone hasn't been showing up for class, often; I mean sometimes it's just a lazy student who can't reach them and that's it. But oftentimes, perhaps even the majority of times, there's usually a reason why someone's not showing up for class. And so you try to seek out this person, and say, what's going on, why is this happening? And other times, you know, a student will just come and say, this is, this is something that's happened. At one point a student told me that a family member just died  and not because they're trying to get out of a homework assignment, but because a family member just died….Likewise, I'm not that perceptive. So I don't want to say, oh, it's, I can read all the students in the room. Or, you know, if anything I'm more clueless than perceptive." ~Logan

This quote illustrates two mechanisms through which faculty obtain information about students. They can either notice academic changes (e.g. missing class or homework) and then proactively reach out to students for contextual information or students can initiate direct communication with the faculty when they experience personal issues.This quote also illustrates that even the most attentive instructors may not be aware of their students' situations.

We find that a set of faculty members feel like students do not communicate with them and that this prohibits them helping even if they want to.

> "One of the main obstacles I see is the students are not very open with faculty members. Maybe obviously because they don't want to discuss their issues that might be going on with the faculty member. If they can discuss openly with a faculty member, then we could, you know, try and help them go through the situations. Yeah, that's the obstacle. I guess the students are not very open to discussing the problem they're having." ~ Kali

Some faculty members address this by intentionally creating opportunities for students to reach out.

> "There is also something of a soft middle road as well that some of us employ. I semi-regularly will include a question on my homeworks or even quizzes along

> the lines of: "How are things going? Is there anything you'd like to tell me about? Feel free to use this space if there's something you'd like to mention. We can discuss it further in person if you wish, or just set up a time to chat if you prefer." That kind of thing seems like it's somewhere in between. ie, it's not that I've noticed a change or something of concern, but it's also not entirely just upon the student as it's an open invitation."~ Logan

We see that faculty still hold students responsible for communicating their struggles but facilitate the communication by showing that they value this information and explicitly asking about it.

We asked faculty to recall a time they noticed a student was struggling outside of the classroom setting and were surprised to find that some could not recall an example. There was a strong correlation between faculty that were open to discussing about themselves and those that stated that students came to them for support/advice when dealing with non-academic issues. 80% of participants who were open about their identities with students were able to recall examples of students coming to them with issues. Five participants could not tell us about a non-academic struggle a student might have faced.

### *4.2  Faculty Personas of Identity Sharing*

We embody faculty responses in four personas. Participants often embody multiple personas, adapting to different contexts as needed. In the following section we delineate the distinctive qualities of each persona and  elucidate their nuanced nature. Finally, we explore the correlation between these personas and students' willingness to bring issues to faculty members. Personas are assigned gender-neutral names and we utilise (they/them) pronouns for all, ensuring inclusivity and neutrality in our characterization.

**Brooke the "Build Trust Sharer"** strategically **shares elements about themselves** to establish a foundation of trust with students, fostering reciprocal sharing. Brooke often worries about students' feelings of safety in the classroom and looks to build space and opportunity for students to bring their whole selves. They take a holistic approach to learning and want students to ask questions both in and out of class. When speaking about their connection with students they say things like,

> "The hope is that I have a good enough relationship with some of the students that they don't shy away from telling me things. And generally, that seems to be true. Because I might know all the ones that have an attention problem or anything else, because I freely admit, during orientation, that I have [mental health] problems. Over the last 20 years, I've gotten far more blatant about it."

For Brooke, having a good relationship with their students is critical. They value knowing personal issues about their students such as mental health or neurodiversity and see this information as instrumental to their teaching practices. Brooke finds sharing their own personal details relevant to the classroom and uses their voice to inspire students to open up about themselves. Brooke is often aware of the power dynamics in the classroom and recognizes their role in trying to alleviate barriers students might face. Brooke is insistent on creating a safe environment where everyone can ask questions and fostering a growth mindset in their students.

Brooke sometimes uses **proxy** narratives to build trust. Since they are aware of how their identity comes into play in learning spaces, they sometimes bring others from diverse backgrounds into their teaching space, for example, involving undergraduate teaching or learning assistants:

> "Although, one thing I am aware of is that even if I don't feel like I'm engaging these identities, I am, necessarily just by being in the space. And so I want people from different backgrounds to feel comfortable asking questions. And that is why I really, really liked working with the TAs or the learning assistants. And sometimes I feel like it's better if the learning assistant or the TA is a woman rather than another man, because my hope is that for some of the female students who would maybe feel intimidated by me, that they feel more comfortable talking to both someone closer age up here, and also the woman undergraduate.. woman who's going through similar sort of experience with them. So that is something that I am cognizant of. Where I tried to be maximally welcoming by myself. But I know that that's not always going to hit for everybody, for whatever reason. So if I can have an assistant who represents different life experiences, that's nice, because that covers more backgrounds, and maybe makes some people more comfortable."

Brooke values students being comfortable to ask questions and feel safe. They recognize that their visible identity might be intimidating to some students, so intentionally seek out folks from different backgrounds to help alleviate that potential issue with students.

**Nour the Identity Navigator** positions themselves as a guide, assisting students in navigating their identities throughout college. They may assume a parental role or act as a role model, believing it is essential to discuss their own identities for the benefit of students. Like Brooke, Nour acknowledges the importance of identity within their profession, but also reflects deeply about their identity and those of their students as they navigate the college space. They see themselves as a mentor in and outside of the classroom to students who might not even be in their classes. They want to help students grow into themselves and think critically about their place within society.

Nour sometimes takes a **role model approach** or a **parental one**. Nour discusses the importance of identity sharing in their classroom while taking a role model approach, they say,

" I will say whatever prompts I'm giving them [about student identity], and they can share as much or as little about them as they want…. But I told them that okay, I will first answer the prompts before they do it. So that, you know, I have to lead by practising what I'm asking them to do. So then I share things about me, you know, my identity and other things."

We see Nour role modelling the identity assignment, setting up the norm for future classes, and making their identity open to all students. Other times Nour takes on more of a parental approach, as illustrated below with Nour talking about research students.

" So when I think about students, you know, I try to remember that, like, all the people I'm interacting with, they're like, someone's mom or dad or their daughter or son or child, you know, and that helps humanise them. Oftentimes I've thought like, Alright if my own kids were going here, what would I want this place to be? And that kind of makes it real? Because it's like, I wouldn't want my kids to have to put up with that [an issue]. I mean, granted, like, that might be kind of normal, but doesn't mean it's good. And I think we can do better."

They talk about how they think of their own kids when they interact with students. These actions allow them to humanise the students and recognize that they are more than just students, they are people navigating their own identity and family dynamic. In this quote Nour also illustrates that they do not want to stick to the status quo, and want to make things better for students. Nour often talks about building a safe community, similar to Brooke, as an important teaching practice but also highlights larger-scale cultural issues that students will have to navigate either in college or in their lives after graduating. Faculty embodying the Nour persona were the only ones to discuss theme 4 of the teaching values, which centred on topics like fostering students' awareness of their societal roles and encouraging them to contemplate their personal ethics and community impact while paying attention and acknowledging the interaction with their identities.

**Casey**, embodying the "**Cautious Sharer**" persona, is willing to share identity-related information, yet exercises caution due to apprehensions about causing harm or fostering distance between themselves and students. When analysing the data two distinct facets emerge: being cautious for the sake of the students and/or being wary of potential backlash to self. Casey also worries about creating a safe and inclusive environment for their students and does not want to do anything that could potentially jeopardise students' sense of safety.

Casey can be cautious **for their students** or **worried for their own sake** when talking about different parts of their identity. Thinking about talking about religion in a classroom setting is one of the main issues Casey faces, as illustrated by the following quote. "Religion is not something I address in the class at all, just it's not the subject of the class. And then employment-wise, I don't think I should be talking about my religion in class". They say it wouldn't be smart for their career to speak about religion. Casey sees religion as a sensitive part of their identity and thus chooses not to talk about it. Casey also views religion as a tricky topic when trying to connect with students.

> "I think there's some things I could say that would encourage a subset of students to be like, oh, yeah, I can kind of connect with that. And there'd be like, other people that will be like, Wow, now I just realised how much more different than you are, you are than me…things like religion and family have always been compartmentalised… like in the sense that you're on campus and there's just not that much space to talk about those things."

Here they wonder about the impact self-disclosure will have on students' ability to relate to them. They see religion as an important part of their identity but don't feel like there is space for it on campus.

Furthermore, Casey can also be cautious about making jokes in class and the backlash that might have on how students see them in relation to their identity and the cultural norms of physics, stating:

> "It's mostly on us- on the faculty, because we are expected to figure out where the students are comfortable, and then allow something like a culture to play a role. But it doesn't really happen with the students in class. So things that are cultural, that's why we don't come out funny, like a lot of professors, because we have to figure out which one is funny in this culture versus our culture. And also, there is a gender issue if women are funnier in class, they sound goofier. And they, the students, lose respect."

We note that the data informing Casey's persona predominantly originates from white faculty members. A plausible inference is that acknowledging one's privilege fosters an awareness that sharing aspects of an identity aligned with the 'normal majority' can marginalise those lacking such privilege.

Finally, **Wray**, with a "**Walled Off**" approach**,** consciously refrains from sharing personal details within professional settings, maintaining a distinct separation between private life and interactions within the classroom or with fellow faculty members. Wray is motivated by two factors that sometimes intersect. They feel their personal life is **not relevant** to their professional activities and value the affordances that come along with having a professional identity for the workplace. Sometimes Wray also had **negative past experiences** with sharing that prompted them to now keep these aspects separate. This could be in the form of feeling uncomfortable when boundaries get crossed or facing external repercussions from choosing to share.

An example of intrinsically valuing the separation of personal and professional lives not wanting to get involved in students' lives, Wray says, *"So usually, I didn't go too personal with my students..But I don't.. like I said, I didn't get too involved in those*

*[students' personal issues]. Just do our standard policy".* Wray often refers to "standard policy" to decide what to do when faced with student issues. They try to not get too involved as they don't see it as their place as an instructor. On the other hand, Wray can also be motivated to stay distant by past experiences.

> "Those things [social research group events] get very firewall for me. I don't. I have had parties to celebrate, you know, people were graduating or whatever at my house. I would say in general, that makes me very uncomfortable when I do. I don't do it anymore"

This quote illustrated how Wray tried something in the past and then realised they were not comfortable with that experience (in this case celebrating with students) so decided to not continue doing it moving forward. While they do not value sharing their personal lives in a professional setting, Wray cares about using effective teaching practices. Wray thinks the most important qualities of teaching are "being well prepared" and "making physics interesting," which were sorted and themed under theme three, teaching practices. When thinking of teaching Wray cares about putting a personal effort into the subject and trying to keep students' attention.

Table 2 summarises the key characteristics of the four personas, including how they share identity and their particular teaching values.

Table 2.Summary of Faculty Personas.

| **Brooke**<br>**Trust Builder** | **Wray**<br>**Walled off** | **Casey**<br>**Cautious sharer** | **Nour**<br>**Identity Navigator** |
|---|---|---|---|
| **Facets**<ul><li>Using Self</li><li>Using Proxy</li></ul> | **Facets**<ul><li>By Value</li><li>By Experience</li></ul> | **Facets**<ul><li>For Self</li><li>For Students</li></ul> | **Facets**<ul><li>Role-Model</li><li>Parental</li></ul> |
| **Identity Sharing**<ul><li>Intentionally shares in order to build trust with others.</li><li>Recognizes the role Identity plays in the classroom</li><li>Create a classroom norm of open communication.</li></ul> | **Identity Sharing**<ul><li>Intentionally does not share with others.</li><li>Values a distinct disconnect between personal and professional life.</li><li>Create a classroom environment where students can focus on the material.</li></ul> | **Identity Sharing**<ul><li>Carefully decides what to share and what not to share about themselves.</li><li>Worried of possibly alienating students or themselves.</li><li>Create a classroom environment that minimises tensions.</li></ul> | **Identity Sharing**<ul><li>Intentionally shares about themselves in order to help students navigate the challenges of growing up.</li><li>Recognizes the role identity and social issues play on learning.</li><li>Create a classroom environment focused on personal growth.</li></ul> |
| **Teaching Values**<ul><li>Values creating safe spaces for students to be able to participate.</li><li>Values utilising active learning methods.</li></ul> | **Teaching Values**<ul><li>Values utilising active learning methods.</li></ul> | **Teaching Values**<ul><li>Values utilising active learning methods.</li></ul> | **Teaching Values**<ul><li>Regards teaching students about the intersection of society and self as key to their teaching practices.</li></ul> |

### *4.3 Distribution of Personas*

Personas do not map directly onto interview participants, but rather categorise motivation and decision-making in a way that depends heavily on context. The number of participants contributing to the development of each persona is depicted in Figure 2. Specifically, Brooke was represented in 13 interviews, Wray in 6, Casey in 10, and Nour in 8. This distribution highlights that Brooke and Casey were the most frequently coded personas within our dataset. Most participants displayed more than one persona when talking about their experiences and how they chose to interact. Figure 2 also shows which personas arise in each participant, with 13/19 participants displaying aspects of at least two personas.

*Figure 2.* Distribution of personas in participants

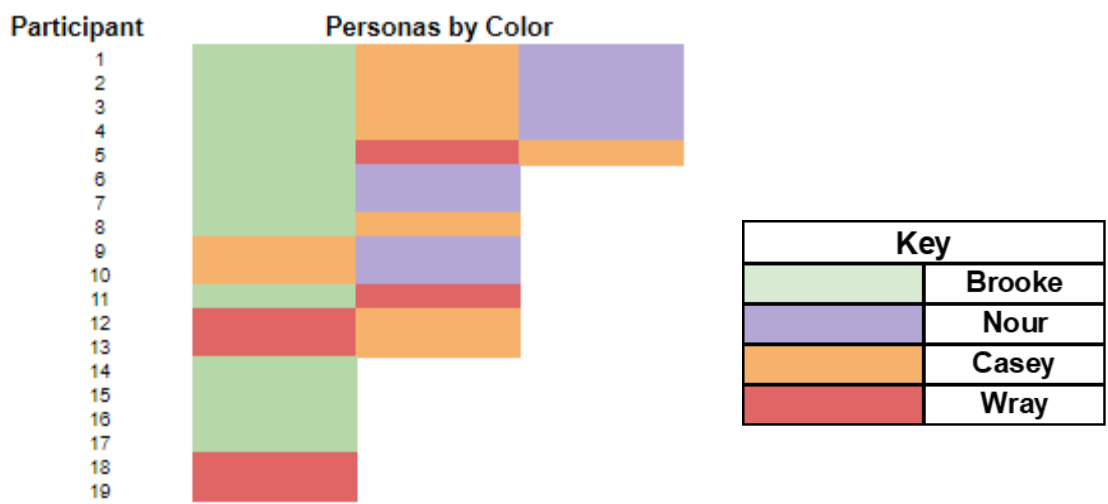

Coding of individual participants by persona. The most common combination of persona was a mix of Brooke, Casey, and Nour, which is examined through the example of Joe in section 4.3. Six of the participants were coded for only one persona.

Faculty manifestations of personas is often context-dependent. For example, one participant, Joe, shows aspects of three distinct personas in interviews. Embodying aspects of Brooke, Casey, and Nour in different contexts. Figure 3 shows a synopsis of Joe's depiction of his three different personas and the specific contexts in which each appears. In the classroom, Joe primarily adopts the personas of Brooke, intentionally sharing about himself to help students trust him, and Casey, exercising caution when discussing concealed identities such as religion. Conversely, in a research mentor setting, Joe exhibits traits of Nour, engaging in discussions about identity within the physics culture while collaborating with his students.

Figure 3. Joe as an example of the context-dependent nature of personas

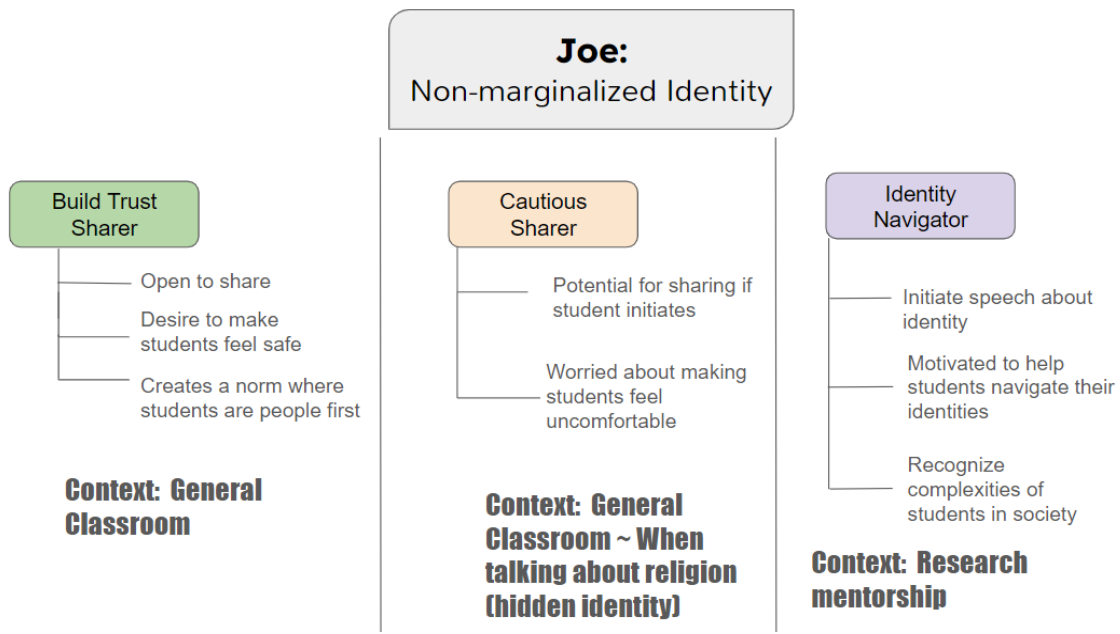

This figure outlines the contexts in which Joe embodies his three personas (Brooke, Casey, and Nour). In the context of the classroom Joe acts as a build trust sharer, wanting to create a safe environment for his students to bring up issues they are facing both in and outside of the classroom. He is cautious when it comes to speaking about his religion as he worries it might alienate some students. He embodies Nour in the context of taking on research students, and in one one-on-one setting where he places emphasis on students' identities and reflects on how best to help them navigate.

# 5  Discussion

## 5.1  *Fluidity of Personas*

Personas are dynamic constructs, and faculty transition between personas depending on context, mirroring the dynamic nature of identity, influenced by time, experiences, external recognition, and personal emotions (Avraamidou, 2019; Calabrese Barton et al., 2013; Carlone & Johnson, 2007). It is plausible to also assume that faculty manifestations of personas evolve in time as faculty gain experience in teaching and mentoring and interact with more students. Faculty may become more aware of student circumstances and notice issues that might previously have gone unseen. As faculty grow, they may also become more or less comfortable sharing.

## 5.2     *Personas and teaching values*

We identified four primary themes pertaining to teaching values that informed our personas: 1) fostering a growth mindset, 2) establishing safe spaces, 3) refining teaching practices, and 4) understanding socio-cultural constructs. All four personas emphasise the importance of enhancing their teaching practices, and the concept of a growth mindset was common across all personas, aligning with its widely recognized significance in effective learning. We originally expected that the majority of participants will discuss (3), as it pertains to individual teaching practices, such as being well-prepared for class, a trend evident in our data. However, only two participants deviated from this expectation. One delved into discussions about students rather than focusing on his practices, while the other briefly touched upon the topic before transitioning to mentorship. This is not too surprising given the semi-structured nature of the interview protocol and thus not a meaningful anomaly. Creating safe spaces was a common theme among all personas except for Wray (the Walled off) and particularly prevalent in Brooke (Build trust). Most intriguing, the exploration of socio-cultural constructs was evident only in Nour (identity navigator). Faculty who share their identity in order to assist students in navigating their own identities also value integration of students and physics within a broader socio-cultural context. Nour embodies a goal for conveying not only subject matter but also societal implications. Consequently, faculty members embodying characteristics akin to the Nour persona can play a crucial role in integrating social justice principles into teaching and mentoring practices.

## 5.3     *Persona and Empathy*

Faculty members often lack training in effective teaching methods, incorporating research-based strategies, and navigating best practices for interacting with students and colleagues from diverse backgrounds. Research has consistently shown that students leave STEM fields due to the practices and attitudes of faculty, and yet noticing issues (e.g. of racism, sexism, or othering) is a persistent challenge. Nonetheless, faculty members do express concern for their students' well-being, and an important step toward understanding student context is open conversation on both parts. We find that faculty engagement in identity sharing often correlates with students sharing their experiences as well, and 80% of faculty interviewed who were explicitly open about their identities reported students coming to them with personal issues. On the contrary, four out of the six participants used in developing Wray's persona reported not recollecting occasions when students approached them to discuss non-academic matters. The one faculty that we did not code for Wray but also did not communicate an example of student non-academic struggle, might have chosen not to discuss things to avoid being identified. The interviewer (AH) also did not have a meaningful insider connection to the participant so they might not have felt safe or wanted to take on an increased mental load to share. We recognized that the interview protocol is very personal in nature and thus building and sustaining a mutual relationship with participants is instrumental to reflecting faculty members' thoughts in a grounded manner, which we did not have with all participants.

      Our research reveals that even well-intentioned faculty members often rely on students to come to them with issues they face. However, this places an undue burden on students, particularly those who are most vulnerable to discrimination, as they may hesitate to speak up (McLeod, 2011). Faculty members, assuming that students will raise concerns, overlook the hidden curriculum embedded in higher education, as not all students know what to do when issues arise in the classroom or outside. Marginalised

students often perceive greater power barriers and have fewer resources to know the social structures within academia (Nasir & Cobb, 2007). Since faculty tend to gain most of their contextual information through students direct communication, it is imperative to develop relationships where that is possible. Incorporating check-in type questions within the coursework(homework, quizzes, discussions) was only mentioned by one participant but could be a mechanism to create space that lowers the initial hesitancy of students reaching out. The personas outlined show how identity-sharing (an action taken by the faculty) can enhance student communication, ultimately creating opportunity for empathetic actions.

### *5.4      Limitation*

There are several limitations in the design of this study and in using personas as an analytical methodology. We acknowledge that our findings are limited to what faculty chose to share with the research team. Though we tried to steer interviews away from specific names and times, faculty sometimes indicated that they felt uncomfortable disclosing information that could be identifiable. Getting rich data about faculty-faculty interactions requires building a level of trust between the interviewer and participants, as such most of our results and conclusions centre on student-faculty interactions.

In constructing the persona there were often overlaps in faculty's views of themselves and their teaching values, so choices had to be made in how to categorise teaching values in ways that did not focus solely on pedagogical tools. Interviews were not centred on faculty's identity-sharing practices, so interview data was sparse and often interrelated to other topics; however, most participants were asked at least one question directly related to their identity-sharing as a faculty member. This was in the context of either the classroom or research group or department. On the other hand, discussions of motivation were more ambiguous, with fewer faculty explicit about why they chose to share or not share parts of their identity.

We also only interviewed faculty within physics at a single R-2 institute within the United States, so our four personas reflect the characteristics and context found within our sample. Personas are also fluid in time so instructors who might fit into one persona at this moment might develop into another. We tried to sample faculty from different times in their careers; however, the data was more highly weighted towards early-career folks. Due to the scope of this study, we did not interview community college instructors or faculty in different STEM disciplines which would be interesting for further work.

Our findings are also limited to physics faculty from a particular department. Physics, like any academic discipline, has a unique culture developed through a hidden curriculum. This hidden curriculum consists of epistemology (how we decide something is true), ontology (how we categories the observable world), and discourse (how we communicate to understand and develop new knowledge), and has been shown to be different in physics compared to other fields such as maths and biology (Redish et al., 2010; Wong et al., 2023) . Studies looking directly at the discipline have found that physics is strongly aligned with notions of intelligence, masculinity, and a fixed mindset of physics being an inherent talent instead of a body of knowledge that is learned through multiple steps and struggles (Archer et al., 2017, 2020). These differences indicate that physics faculty might be unique compared to the STEM faculty and thus might develop dissimilar practices around identity sharing and empathy development. All participants also belonged to the same department, which contains its own norms and expectations that might differ from external departments. These differences might limit or enable different forms of communication--in particular, identity sharing--both within and outside of the classroom. Departments create their own micro-culture, where faculty communicate about their practices and gain inspiration from one another. The high prevalence of Brooke as a persona might be an indication of the department seeing being open as a norm. This particular department is also heavily engaged in creating active learning spaces and has transformed the introductory physics classes to use a workshop style, which is not representative of most physics departments throughout the US.

# 6    Conclusion

This research provides insight into physics faculty's attitudes towards sharing information about themselves within differing contexts. In deciding what personal information to share with students, instructors navigate an interwoven road of pedagogical changes and DEISJA efforts. Our findings indicate that faculty rely on distinct motives when determining how and when to share aspects of their identities. These motives (personas) are fluid, with faculty moving between different personas depending on context. We found four distinct personas Brooke, Wray, Casey, and Nour and illustrated similarities and differences between them. Brooke, known as the Trust Builder, prioritised creating an environment of trust by openly discussing their identity, aiming to foster student openness. Wray, adopting a Walled-Off approach, separated personal and professional life due to past negative experiences or a belief in the importance of that division. Casey, embodying the role of the Cautious Sharer, expressed concerns about potential alienation or backlash, thus decided to not share elements of their identity. Nour, characterised as the Identity Navigator, shared personal experiences to assist others in navigating their own identities, acknowledging the interaction between self, culture, and society.

The personas outlined in our study serve to elucidate student-faculty interactions as a didactic process reliant on two-way communication. Each persona embodies specific values guiding considerations such as appropriate discourse, teaching priorities, fairness, and support for students in need. Obtaining contextual information is instrumental to the development of empathy with all personas indicating they mainly rely on students approaching them with their issues, creating a potential barrier to faculty empathy and in retrospect meaningful action. Given the diversity in faculty identity sharing, it is reasonable to expect variations in how empathy is developed and perceived within higher education. Further studies need to look at students' perceptions of faculty roles and empathy. It is important to build an understanding of what factors influence what and why students decide to share with their faculty, and how much goes unsaid. It is equally as important to investigate faculty members' conceptualizations of empathy and the factors shaping their perspectives, as well as what factors influence their decision making. With more information faculty can do a better job at perspective taking and imagining what it might be like in their students' shoes. Pop culture likes to throw around the word empathy as the perfect answer to building a kinder, more just world; however, as Dylan alluded to, developing empathy is hard and requires open communication from all parties. This is often difficult to do given the hierarchical nature within academia. We believe this work is a first step towards understanding practical steps that can be taken to help elucidate empathy in the classroom. Most critically, our results point towards the interaction between identity sharing, gaining student information and the development of empathy, as well as highlighting the complexities involved in identity sharing. While the idea of empathy has gained traction in higher education, without further exploration of the steps and factors contributing to empathy development, it becomes challenging to devise and apply practical tools effectively for leveraging empathy as a catalyst for fostering change.

# 7 Declarations

### 7.1. Availability of Data and Materials

The datasets generated and/or analysed during the current study are not publicly available to preserve the confidentiality and anonymity of the participants but are available from the corresponding author on reasonable request.

### 7.2 Acknowledgments (include funding and disclosure statement)

This work was supported by the NSF under Grant [#2222337]. We would also like to thank the CASTLE members for their insight and ideas. We appreciate the RIT Capstone initiative which allowed for meaningful undergraduate research. *The authors report there are no competing interests to declare*.

### 7.3 Authors Contributions

AH: Conceptualization, data curation and collection, formal analysis, investigation, methodology, project administration, resources, supervision, writing—original draft, writing—review and editing, visualisation.

AB: Conceptualization, data curation and collection, investigation, methodology, project administration, validation.

DN: Conceptualization, resources, supervision, funding, writing—review and editing, visualisation.

SF: Conceptualization, resources, project administration, supervision, funding, writing—review and editing, visualisation.